\documentclass[prl,aps,twocolumn,showpacs]{revtex4}

\usepackage{subfigure}
\usepackage{color,amsmath,graphicx,epsfig,amssymb}
\usepackage[T1]{fontenc}
\usepackage{pslatex}
\usepackage{txfonts}

\begin{document}

\def\pd#1#2{\frac{\partial #1}{\partial #2}}
\def\bk{{\bf k}}
\def\bx{{\bf x}}
\def\grad{{\mbox{\boldmath$\nabla$\unboldmath}}}

\title{\bf Approximating Steady States in Equilibrium and Nonequilibrium
Condensates}
\author {Hayder Salman}
\affiliation {School of Mathematics, University of East Anglia, Norwich Research Park, Norwich, NR4 7TJ, UK}

\begin {abstract}
We obtain approximations for the time-independent Gross-Pitaevskii
(GP) and complex GP equation in two and three spatial dimensions by generalizing the divergence-free WKB method. The results include an explicit expression of a uniformly valid approximation for the condensate density of an ultracold Bose gas confined in a harmonic trap that extends into the classically forbidden
region. This provides an accurate approximation of the
condensate density that includes healing effects at leading
order that are missing in the widely adopted Thomas-Fermi approximation.
The results presented herein allow us to formulate useful approximations to
a range of experimental systems including the equilibrium properties of a finite
temperature Bose gas and the steady-state properties of a 2D nonequilibrium
condensate. Comparisons between our asymptotic and numerical
results for the conservative and forced-dissipative forms of the GP
equations as applied to these systems show excellent agreement between
the two sets of solutions thereby illustrating the accuracy of these approximations.
\end{abstract}
\maketitle

The complex Gross-Pitaevskii (cGPE) equation, also known as the cubic Ginzburg-Landau equation, or Nonlinear Schr\"{o}dinger equation (NLS) arises
in many branches of physics. It has successfully been used to model phenomena such as nonlinear (optical) waves, second order phase transitions, superconductivity, superfluidity, and Bose-Einstein condensation of atomic gases as well as quasiparticle excitations. In the context of superfluidity, the GP \cite{GP} equation has served as an excellent model for atomic gases while the cGPE has faithfully reproduced a number of key phenomena observed in experiments on nonequilibrium condensates of quasiparticle excitations. A key feature of this equation is that it describes phenomena dominated by different physical processes that lie on either side of a nonlinear turning point. The nonlinear turning point is governed by a second Painlev\'{e} transcendent, which is a canonical equation arising in all of the contexts mentioned above, and more generally in systems that are nonlinear generalizations of an underlying linear problem governed by a second order differential equation. It is, therefore, no surprise to see that it also arises in the nonlinear Landau-Zener problem \cite{Itin}. Yet, unlike Airy's equation, which governs the classical turning points of the linear Schr\"{o}dinger equation, a uniformly valid solution or approximation for this equation has remained a formidable challenge.

Focusing on the problem of Bose-Einstein condensates for ultracold gases, we note that the most commonly used method for determining the steady state solutions of the (GP) equation is based on the Thomas-Fermi approximation. However, in general, it is not possible to extend the Thomas-Fermi (TF) profile uniformly into the classically forbidden region \cite{PitStr} because of the need to solve the  second Painlev\'{e} transcendent \cite{PitStr,Dalfovo96}. The difficulty in obtaining uniformly valid approximations has meant that the TF approximation has been used in many circumstances where it is clearly invalid.  A particularly important example arises in determining the equilibrium properties of a Bose gas at finite temperature. To determine the equilibrium properties of a weakly interacting Bose gas within a confining potential, Nikuni and Griffin \cite{Nikuni01} invoked
a TF profile for the condensate density, and a WKB (equivalently a local density approximation (LDA)) for the thermal excitations. This in turn produced an unrealistic cusp shaped distribution of the thermal cloud density at the edge of the condensate. Since the thermal cloud attains its maximum value at the edge of the condensate, the error introduced by the TF approximation introduces significant errors in the computation of the thermal cloud density. 

Aside from its relevance to experiments, knowledge of the equilibrium properties of the system is also especially important in finite temperature models of Bose gases. A common feature of many of these works (e.g.\ $c$-field methods \cite{cfield}) is to model the macroscopically occupied coherent part of the system using a classical field that is coupled to an incoherent part of the system that is made up of higher energy scarcely occupied modes. An energy cut-off must then be specified that determines which subset of the system is modeled as a classical field. This cut-off is typically determined from the equilibrium properties of the system \cite{burnett} requiring a solution of the Bogoliubov-de-Gennes (BdG) equations in the Bogoliubov or Popov approximation. Being able to specify the cut-off in a simpler way through useful analytical approximations would be particularly useful for such finite temperature models.

In addition to the above problems, there has been a surge of activity in recent years in the properties of nonequilibrium condensates. These can include exciton-polariton \cite{Kasprzak} or magnon \cite{Demokritov} condensates where a condensate is created through coherent pumping to balance the dissipative processes that exist in such systems. The action of these nonconservative effects can significantly alter the form of the condensate from the TF profile as was illustrated by Keeling and Berloff \cite{Keeling08}. Given these recent developments, useful approximations that go beyond the TF approximation are clearly needed.

In \cite{Hyouguchi02}, a method was proposed that resolves the divergences that arise around the turning points in the classical WKB methods. In contrast to other approaches \cite{Witthaut06}, the divergence free WKB method also provides a uniformly valid solution for the ground state wavefunction of the NLS equation. However, the divergence free WKB method has found limited applications partly because it is restricted to 1D systems. Since Bose-Einstein condensation is also studied in systems of higher dimensions, we will begin by extending the results of the divergence free WKB to higher dimensions. For generality, we consider the cGPE given by
\vspace{-0.1cm}
\begin{eqnarray}
i \hbar \partial_t \psi &=& -\frac{\hbar^2}{2m} \nabla^2 \psi + V_{\mathrm{ext}} \psi + g |\psi|^2 \psi + i[\mathcal{S}-\mathcal{D}] \psi,
\label{eqn_GP}
\end{eqnarray}
where $\mathcal{S}$ and $\mathcal{D}$ denote pumping and dissipation respectively.

To begin, we will focus on the $T=0$ Bose gas where $\mathcal{S}=\mathcal{D}=0$. In this case, the equation has two constants of motion corresponding to a fixed total energy which in 3D is given by the Hamiltonian $H=\int (\hbar^2/2m|\grad \psi|^2 + V_{\mathrm{ext}} |\psi|^2 + \frac{g}{2}|\psi|^4) d^3\bx$ and the total number of particles $N=\int |\psi|^2 d^3\bx$. The parameter $g=4\pi a_s\hbar^2/m$, where $a_s$ is the $s$-wave scattering length for the interatomic interaction potential. In this work, we will be concerned with a harmonic trapping potential of the form $V_{\mathrm{ext}} = \frac{m}{2}(\omega_x^2 x^2 + \omega_y^2 y^2 + \omega_z^2 z^2)$. Following \cite{Fetter98}, we will non-dimensionalise Eq.\ (\ref{eqn_GP}) using the energy scale $\hbar \omega_{ho}$ and the length scale $R = a_{ho} \left( 15N_0a_s/a_{ho} \right)^{1/5}$, where $N_0$ are the number of particles in the condensate (note $N_0 \ne N$ for finite temperature models based on the $c$-field approximation), $a_{ho} = \sqrt{\hbar/(m\omega_{ho})}$, and the average oscillator frequency is defined in terms of the three oscillator frequencies as $\omega_{ho} = (\omega_x \omega_y \omega_z)^{1/3}$. 
Letting $\psi \rightarrow \sqrt{N} \psi \text{e}^{-i\mu t}$ leads to
\vspace{-0.5cm}
\begin{eqnarray}
i \partial_t \psi &=& -\frac{\epsilon^2}{2} \nabla^2 \psi + \tilde{V}_{\mathrm{ext}} \psi + \gamma |\psi|^2 \psi - \mu \psi.
\label{eqn_psi_nd}
\end{eqnarray}
where $\tilde{V}_{\mathrm{ext}}({\bf x})=\frac{1}{2}(\lambda_x^2 x^2+\lambda_y^2 y^2+\lambda_z^2 z^2)$, $\gamma = 4\pi N_0 a_s a_{ho}^4/R^2 $, $\epsilon^2 \equiv (a_{ho}/R)^4$ is a small coefficient, $\mu$ is the chemical potential, and the trap anisotropy is given by $\lambda_x = \omega_x/\omega_{ho}$, $\lambda_y = \omega_y/\omega_{ho}$, $\lambda_z = \omega_z/\omega_{ho}$. 
Despite the nonlinearity in our equation, we proceed by seeking variable separable steady-state solutions. Motivated by the divergence free WKB for 1D systems, we express the wavefunction as $\psi(x,y,z) = 
\exp{[(\varphi(x)+\vartheta(y)+\phi(z))/\epsilon]}$. Substituting into Eq.\ (\ref{eqn_psi_nd}), we obtain
\vspace{-0.1cm}
\begin{eqnarray}
0 &=& \frac{-\epsilon}{2} \left( \varphi''+\vartheta''+\phi'' \right)
-\frac{1}{2} \left( \varphi'^2+\vartheta'^2+\phi'^2 \right)
+ \tilde{V}_{\mathrm{ext}}  \nonumber \\ && + \gamma \text{e}^{2(\varphi+\vartheta+\phi)/\epsilon} - \mu. \label{eqn_WKB}
\end{eqnarray}
Differentiating the above equation with respect to $x$, we obtain
\begin{eqnarray}
\!\! 0 = \left[ \frac{-\epsilon \varphi'''}{2}  - \varphi' \varphi'' \right] + \frac{\partial_x \tilde{V}_{\mathrm{ext}}}{\partial x} + 2\gamma \frac{\varphi'}{\epsilon} \text{e}^{2(\varphi+\vartheta+\phi)/\epsilon}. \label{eqn_divfWKB1}
\end{eqnarray}
After eliminating $\varphi''$ with Eq.\ (\ref{eqn_WKB}), we have
\vspace{-0.1cm}
\begin{eqnarray}
\frac{\epsilon^2 \varphi'''}{2} \! = \! \varphi' \left[ \epsilon(\vartheta'' + \phi'') + \varphi'^2 + \vartheta'^2 + \phi'^2
+ 2(\mu-\tilde{V}_{\mathrm{ext}}) \right] + \epsilon \frac{\partial \tilde{V}_{\mathrm{ext}}}{\partial x}. \nonumber 
\end{eqnarray}
If the trapping potential is of the form $\tilde{V}_{\mathrm{ext}}(x,y,z) = \tilde{V}_1(x) + \tilde{V}_2(y) + \tilde{V}_3(z)$ (as for a general harmonic trap), the above equation is variable separable. It then follows that we can write
\vspace{-0.4cm}
\begin{eqnarray}
\frac{\epsilon^2 \varphi'''}{2} &=& \varphi'^3 + \varphi' \left[ 
2(\mu_c-\tilde{V}_1) \right] + \epsilon \partial_x \tilde{V}_1, \label{eqn_varphi} 
\end{eqnarray}
where $\mu_c = \mu + C$. From the boundary conditions $\partial_x \psi = \partial_y \psi = \partial_z \psi = 0$ at $\mathbf{x}=(0,0,0)$ we have $\varphi'(0) = \vartheta'(0) = \phi'(0) = 0$ and so the constant $C=\epsilon(\vartheta''(0)+\phi''(0))/2$. By similar reasoning, one can arrive at analogous equations for $\vartheta'(y)$ and $\phi'(z)$ reducing our 3D problem to a solution of three uncoupled equations. We can now proceed as in the divergence-free WKB to obtain solutions for $\varphi'$ by neglecting the terms of $\mathcal{O}(\epsilon^2)$ on the left-hand side of Eq.\ (\ref{eqn_varphi}) whilst retaining terms of $\mathcal{O}(1)$ and $\mathcal{O}(\epsilon)$ on the right-hand side. As discussed in \cite{Hyouguchi02}, the resulting cubic algebraic equation of the form, $\varphi'^3 + a\varphi'^2 + b\varphi' + c =0$ and similarly for $\vartheta', \phi'$, has three roots one of which corresponds to a uniformly valid solution of the cubic NLS equation.
If $a,b$, and $c$ are real, this root is given by
\begin{eqnarray}
&& \!\!\!\!\!\!\!\!\! \varphi' = (A+B)-\frac{a}{3}, \;\;\; A=-\text{sgn}(R)\left[ |R|+\sqrt{R^2-Q^3} \right]^{1/3},
\nonumber \\
&& \!\!\!\!\!\!\!\!\! B= \left\{ \begin{array}{cc}
Q/A, & (A \ne 0) \\
0 & (A = 0)
\end{array} \right., \;\;\;
Q=\frac{a^2-3b}{9}, \;\; 
R=\frac{2a^3-9ab+27c}{54}. \nonumber
\nonumber 
\end{eqnarray}
The solution we have obtained for $\psi(x,y,z) = \exp{\left[ \int \varphi' dx + \int \vartheta' dy + \int \phi' dz \right]}$, can therefore be evaluated in terms of a quadrature of the integrals in the exponent. This leading order solution includes quantum mechanical effects (e.g.\ healing layers) arising from the kinetic energy terms. Higher order corrections can now be obtained to this zeroth order solution by expanding $\varphi'$ in powers of $\epsilon^2$ such that $\varphi'= \sum_{n=0}^{\infty} \epsilon^{2n} \varphi_n'$. At the next order, we obtain
\begin{eqnarray}
\frac{\epsilon^2}{2} \varphi_0''' = 3 \epsilon^2 \varphi_0'^2 \varphi_1' + \epsilon^2 \varphi_1' [2(\mu_c-\tilde{V}_1)]. 
\end{eqnarray}
Now for a harmonic trap, it turns out that $\varphi_0'$ can be integrated explicitly. Using integration by parts together with Eq.\ (\ref{eqn_varphi}) to express $x$ in terms of $\varphi_0'$, we obtain 
\begin{eqnarray}
&& \!\!\!\!\!\!\!\!\!\!\! \int_0^x \varphi_0' dx' = x \varphi_0' - \int_0^x x' d\varphi_0' =  x \varphi_0' - \frac{\epsilon}{2}\ln |\varphi_0'| + \frac{1}{4}
\left[ \epsilon^2+8\mu_c \varphi_0'^2 \right. \nonumber \\
&& \!\!\!\!\!\!\!\!\!\!\!\ \left. + 4\varphi_0'^4 \right]^{1/2}
 + \frac{\mu_c}{2} \left\{ \ln \left[ 2(\mu_c+\varphi_0'^2) + (\epsilon^2+8\mu_c \varphi_0'^2 + 4\varphi_0'^4)^{1/2} \right] \right\} \nonumber \\
&& \!\!\!\!\!\!\!\!\!\!\!\ -\frac{\epsilon}{4} \text{arctanh} \left\{ (\epsilon^2+4\mu_c \varphi_0'^2)/[\epsilon(\epsilon^2+8\mu_c \varphi_0'^2 + 4\varphi_0'^4)^{1/2}] \right\} - \varphi_0(0). \nonumber
\end{eqnarray}
As far as we are aware, this explicit expression for $\varphi_0'$ provides the first uniformly valid explicit approximation for the condensate density that includes the effects of the healing layers near the edges of the condensate. 
Moreover, this solution applies to 2D or 3D harmonic traps with different oscillator frequencies in each spatial direction allowing cigar and pancake shaped condensates to be obtained. 
At next order, the analytical solution requires the evaluation of a simple quadrature for the integral $\int_0^x \varphi_1' dx'$. The solution, presented above, determines $\varphi$ upto some constant. To specify the constant, we evaluate the value of $\varphi(0)$ from Eq.\ (\ref{eqn_WKB})
It follows that
$\varphi(0) + \vartheta(0) + \phi(0) = \frac{\epsilon}{2} \ln \left\{ \frac{\epsilon[\varphi''(0)+\vartheta''(0)+\phi''(0)]+ 2\mu}{2\gamma} \right\}.$
We note that explicit dependence on the parameter $\gamma$ appears only in the value of $\varphi(0)$. However, its effect is also contained in the chemical potential $\mu$ through the normalization condition $\int |\psi|^2 d^3\mathbf{x} = 1$.

\begin{figure}[t!]
\centering
 \begin{minipage}[b]{0.45\textwidth}
    \centering
    \subfigure[\label{fig_T0} ]{
      \label{fig:mini:subfig:a}
      \epsfig{file=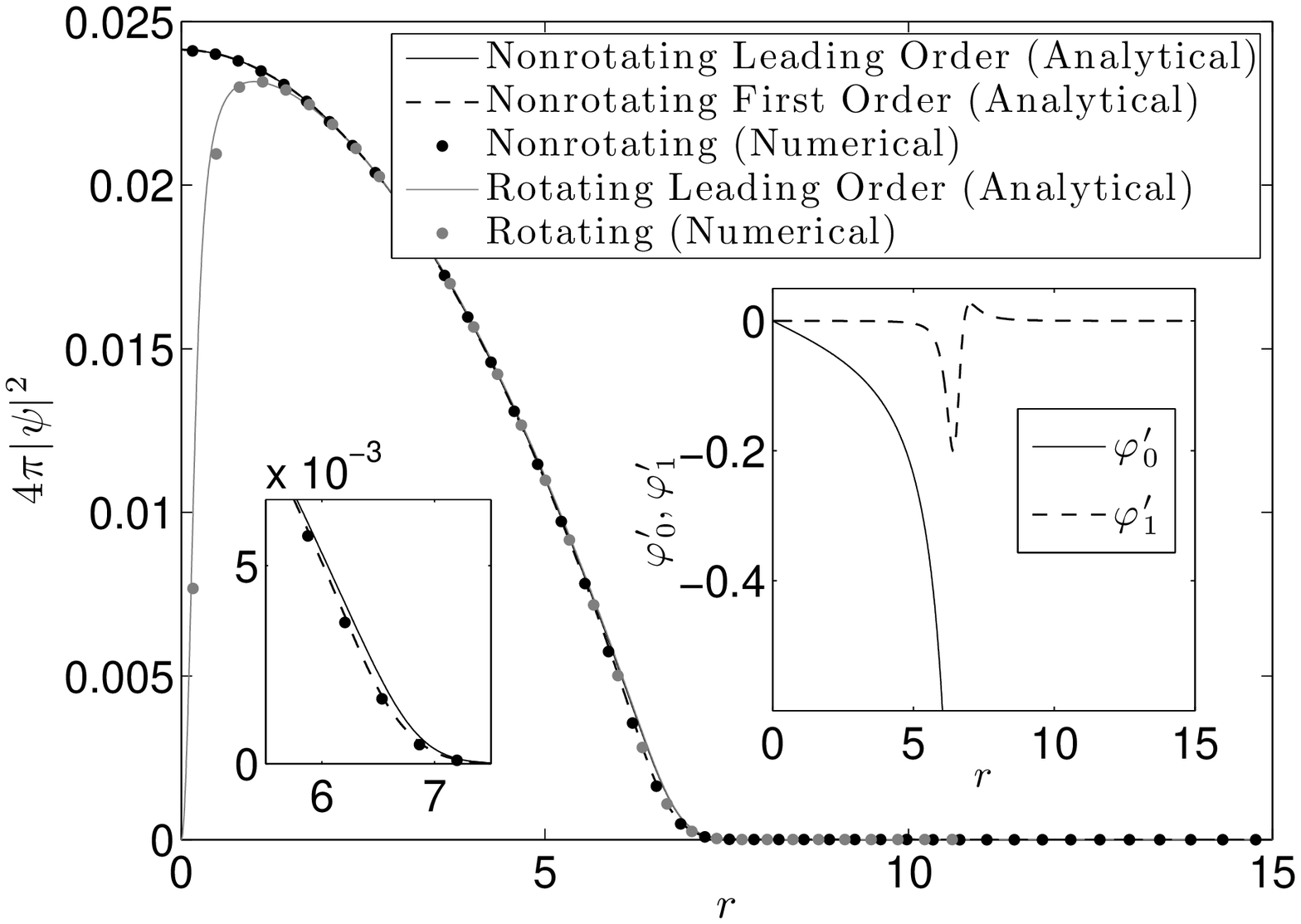,scale=0.45,angle=0.0}}
  \end{minipage}
 \begin{minipage}[b]{0.45\textwidth}
    \centering
    \subfigure[\label{fig_BE} ]{
      \label{fig:mini:subfig:a}
      \epsfig{file=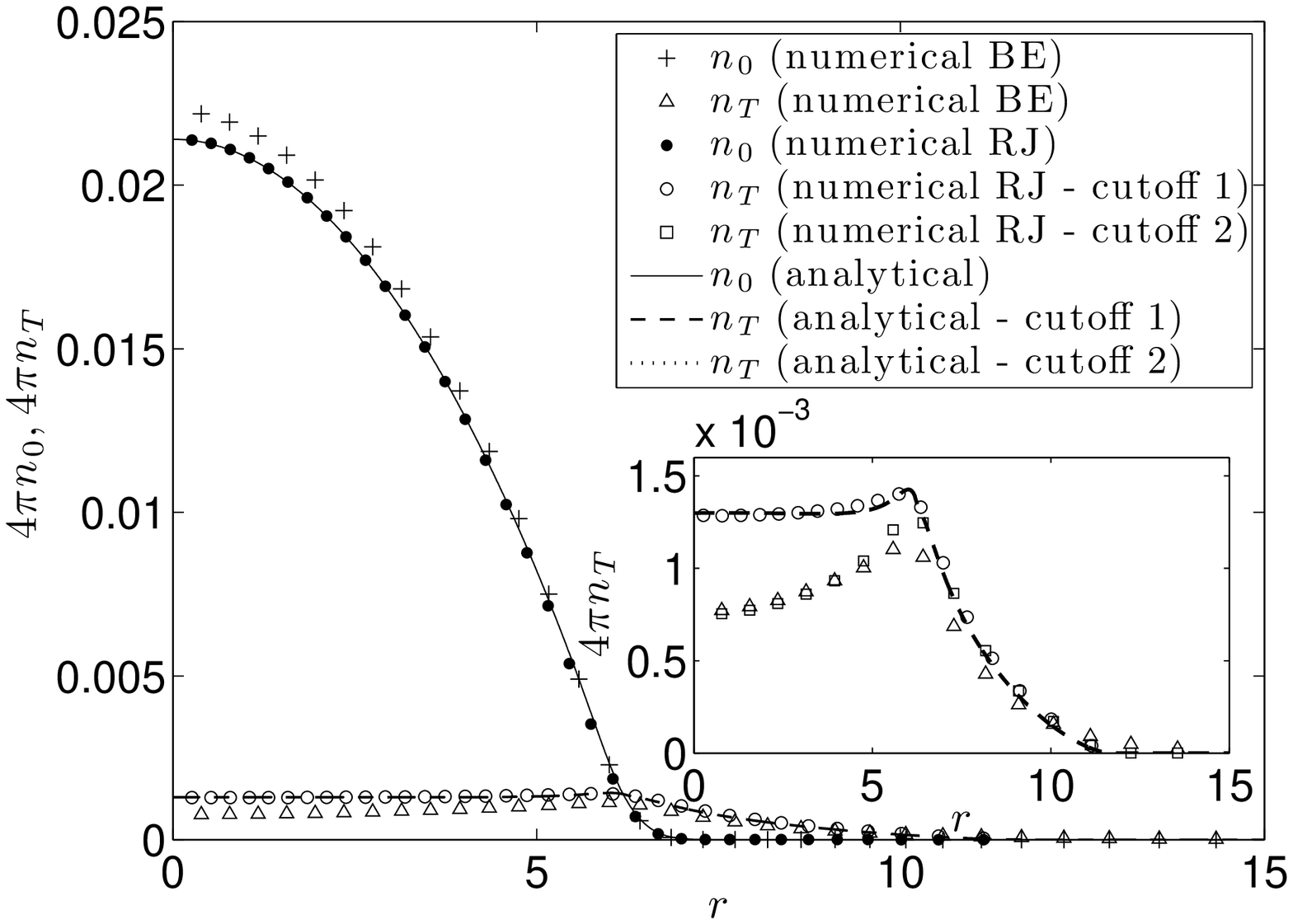,scale=0.45,angle=0.0}}
  \end{minipage}
  \begin{minipage}[b]{0.45\textwidth}
    \centering
    \subfigure[\label{fig_EP} ]{
      \label{fig:mini:subfig:b}
      \hspace{0.2cm}
      \epsfig{file=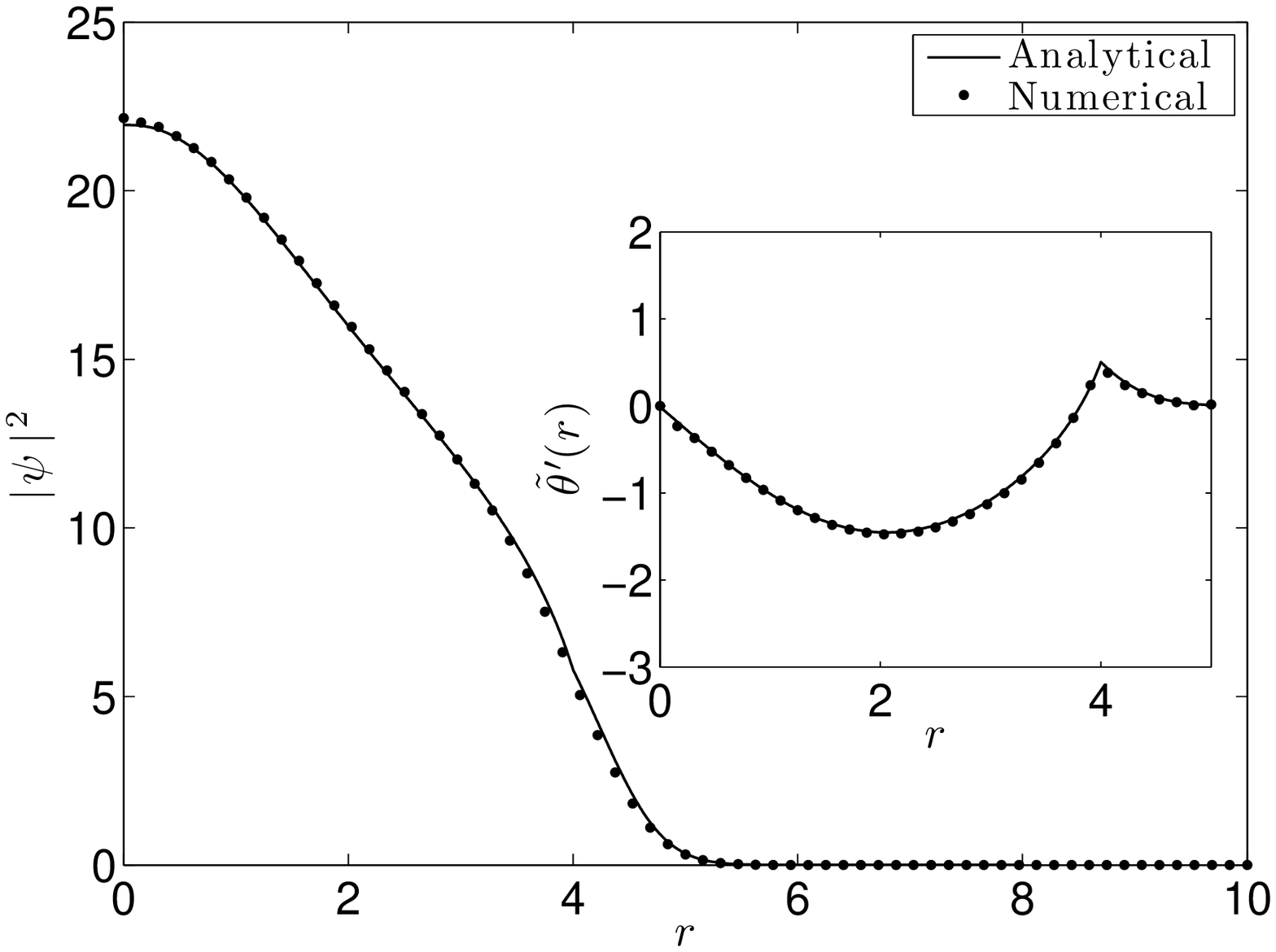,scale=0.46,angle=0.0}}
  \end{minipage}
  \vspace{-0.3cm}
  \caption{Numerical and analytical results ($r$ measured in units of $a_{ho}$): (a) Numerical and analytical solutions of BE condensate at T=0; left and right insets show healing layer, and profiles for $\varphi_0', \varphi_1'$ respectively;
(b) Condensate/thermal cloud densities at T=80nK obtained from numerical and analytical solutions of BdG equations with BE and RJ distributions; (c) Leading order approximation for density of 2D exciton-polariton condensate; inset showing velocity within the condensate. \label{fig_cond}}
\vspace{-0.4cm}
\end{figure}

Figure \ref{fig_cond}a presents a comparison of the condensate density for a spherically symmetric harmonic trap against numerical simulations for the ground state of the GP equation for $\mu=23.05$. By specifying $\mu$ to correspond to the value in our numerical simulations, our asymptotic solutions will approximately satisfy the normalization condition on the wavefunction.
The numerical simulations were performed using Laguerre polynomials  as the basis functions. We see that even at leading order, the results are in excellent agreement with a fidelity of 99.957\%. The small discrepancies that arise can be reduced if we include the contribution $\varphi_1'$ arising from the next order in our expansion as seen in the inset. The numerical and analytical results almost fully coincide at this order of approximation with a fidelity of 99.996\%. The second inset shows the functions $\varphi_0'$, and $\varphi_1'$ illustrating that higher order corrections are localized around the nonlinear turning point.
 
We have also extended the solution to a rotating condensate with a vortex located at the center of the trap by seeking radially symmetric solution. Now following \cite{Langer37}, we first transform the radial NLS equation
into a form suitable for a WKB approximation by setting $\psi(r,z) = U(x,z)$ where $r=\mathrm{e}^x$.
We then seek a variable separable solution of the form $\psi(r,z) = U(x,z) = \exp([\varphi(x)+\phi(z)]+is\theta)$ where $\theta$ is the polar angle. Following a similar procedure as discussed above, we can reduce the problem to two equations for $\varphi'$ and $\phi'$. The equation governing the radial wavefunction for the case $s=\pm1$ is now given by
\begin{eqnarray}
\varphi'^3 + \epsilon \varphi'^2 + (2\mu \mathrm{e}^{2x}-\mathrm{e}^{4x}-\epsilon^2)\varphi' + \epsilon \mathrm{e}^{4x} - \epsilon^3 = \epsilon^2 \left( \frac{\varphi'''}{2} - \varphi'' \right). \nonumber
\end{eqnarray}
In order to truncate the above equation, we perform a local analysis around the origin in which the terms containing the exponentials become exponentially small. Proceeding with this simplification and neglecting the terms on the right hand side, we obtain $\varphi'^3 + \epsilon \varphi'^2 -\epsilon^2 \varphi' - \epsilon^3 = 0.$ This equation has two solutions given by $\varphi' = \pm \epsilon$ which can also be expressed in terms of $\tilde{\varphi}'(r) =\varphi'(x)/r= \pm \epsilon/r$. The solution near the origin is therefore given by $U(x) = \tilde{U}(r) = e^{\int^r \pm \epsilon/r' dr'} = e^{\pm \epsilon \ln r} = r^{\pm \epsilon}$. On physical grounds, we obtain $\tilde{U}(r) = r^{\epsilon}$ as expected. Moreover, if we now substitute this solution back into the right hand-side of the full equation given above, we find that these terms vanish near the origin. This justifies why we can neglect these $\mathcal{O}(\epsilon^2)$ terms whilst retaining the $\mathcal{O}(\epsilon^2)$ and $\mathcal{O}(\epsilon^3)$ on the left hand side which allows us to satisfy the boundary conditions for the vortex at the origin. We, therefore, compute $\varphi'$ by neglecting only the right-hand side of the above equation.

Proceeding as before, we can then obtain an explicit expression for $\varphi$ and $\phi$. Whilst $\phi$ has the same form as the expression given above, $\tilde{\varphi}(r)$ is now given by
\begin{eqnarray}
&& \!\!\!\!\!\!\!\!\! \tilde{\varphi}(r) = r \ln(r) \tilde{\varphi}' + \frac{r \tilde{\varphi}'}{2} - \frac{1}{2} \epsilon \ln(|r \tilde{\varphi}'+\epsilon|) \nonumber \\
&& \!\!\!\!\!\!\!\!\! - \frac{1}{2}
\ln \left( \left| \frac{\mu r \tilde{\varphi}' - \sqrt{(r \tilde{\varphi}')^4 + (\mu^2 - 2\epsilon^2) (r \tilde{\varphi}')^2 + \epsilon^4 }}{r \tilde{\varphi}'-\epsilon} \right| \right) r \tilde{\varphi}' \nonumber \\
&& \!\!\!\!\!\!\!\!\! - \frac{\epsilon}{2} \mathcal{R}\left\{ \arctan \left( \frac{\mu[(r \tilde{\varphi}')^2+\epsilon^2]}{2\epsilon\sqrt{(r \tilde{\varphi}')^4 + (\mu^2 - 2\epsilon^2) (r \tilde{\varphi}')^2 + \epsilon^4 }} \right) \right\} \nonumber \\
&& \!\!\!\!\!\!\!\!\! +\frac{\mu}{4} \ln \left( \left| \frac{\mu^2}{2} - \epsilon^2 + (r \tilde{\varphi}')^2 + \sqrt{(r \tilde{\varphi}')^4 + (\mu^2-2\epsilon^2) (r \tilde{\varphi}')^2 + \epsilon^4} \right| \right), \nonumber
\end{eqnarray}
where $\mathcal{R}(\cdot)$ denotes the real part. The above expression for $\tilde{\varphi}(r)$ is known upto some arbitrary constant that sets the overall normalization condition for the wavefunction. 
In this example, we have computed this constant such that the maximum value of  $|\psi|$ corresponds to the maximum value obtained numerically. We have included a comparison of the computed and analytical profiles in Fig.\ \ref{fig_cond}a. As before, we see excellent agreement even in this case of a rotating condensate with a quantized vortex located at the center of the trap.  

Having illustrated our approximation of the condensate at $T=0$, we now extend the results to a Bose gas at finite temperature. We recall that the BdG system in the Bogoliubov approximation can be written as \cite{Giorgini96,Pethick02}
\vspace{-0.15cm}
\begin{eqnarray}
\frac{-\epsilon^2}{2} \nabla^2 \Phi({\bf x})+ (\tilde{V}_{\mathrm{ext}} + \gamma n_0 ) \Phi({\bf x}) \!\!\! &=& \!\!\! \mu \Phi({\bf x}), \nonumber \\
\frac{-\delta^2}{2} \nabla^2 u_i({\bf x}) + (\tilde{V}_{\mathrm{ext}} + 2\gamma n_0  - \tilde{\mu}) u_i({\bf x}) - \gamma n_0 v_i({\bf x}) \!\!\! &=& \!\!\! E_i u_i({\bf x}), \nonumber \\
\frac{-\delta^2}{2} \nabla^2 v_i({\bf x}) + (\tilde{V}_{\mathrm{ext}} + 2\gamma n_0 - \tilde{\mu}) v_i({\bf x}) - \gamma n_0 u_i({\bf x}) \!\!\! &=& \!\!\! -E_i v_i({\bf x}), \nonumber 
\end{eqnarray}
where $\tilde{\mu} = \mu\hbar\omega_{ho}/k_BT$, $n_0({\bf x}) = N_0|\Phi({\bf x})|^2$ is the density of the condensate, and $n_T({\bf x}) = \sum_{i \ne 0} N_i (|u_i({\bf x})|^2+|v_i({\bf x})|^2)$ is the density of non-condensed particles. These equations are solved subject to the conditions $N = N_0 + \sum_{i \ne 0} N_i$ where $N_i = \{ \exp \left[ (E_i-\tilde{\mu}) \right] - 1 \}^{-1}$ for $i \ne 0$. We have non-dimensionalised our equation for the condensate $\Phi$ as before for the $T=0$ case. The equations for the Bogoliubov modes, $u_i$ and $v_i$ have been non-dimensionalised using the length scale $R$ and the energy scale $k_B T$ which gives $\delta \equiv (a_{ho}^2 \lambda_T^2/R^4)$ with the thermal de Broglie wavelength given by $\lambda_T=(\hbar/\sqrt{mk_B T})$. 

Now for a given experimental system consisting of $N=250000$ $^{87}Rb$ atoms, we have $m(^{87}$Rb$)=1.44$ kg $\times 10^{-25}$ kg, $\omega_{ho}/(2\pi)=50$ Hz, and an $s$-wave scattering length of $a \simeq 5.82 \times 10^{-9}$ m. For typical temperatures of $T=80$ nK, we estimate $\lambda_T/a_{ho} \sim 0.173 \ll R/a_{ho} \sim 6.37$ for a condensate fraction $N_0/N \sim 0.731$. Thus, for typical parameter regimes, $\delta^2 \sim \epsilon^2 \ll 1$ and the condensate varies on length scales much larger than the length scale associated with our excitations. We can, therefore, seek solutions in the form $u_i = \exp{[(\varphi_e+\vartheta_e+\phi_e)/\delta]}$ and similarly for $v_i$ by using a conventional quadratic WKB for the excitations. Combining this with a cubic WKB approximation for the condensate allows us to correctly resolve the nonlinear turning point where the thermal cloud density attains its maximum value.

Motivated by the need to determine the equilibrium thermodynamic properties of a $c$-field simulation \cite{Goral02}, we apply our theory to the BdG system with a Rayleigh-Jeans (RJ) distribution for $n_T$. In the semiclassical approximation, the thermal population with a RJ distribution can be computed exactly as 
\begin{eqnarray}
\!\!\!\!\!\!\!\!\!\!\!\!\!\!\!\!\!\!\!\!\!\! n_T \!\!\! &=& \!\!\! \frac{4\pi k_B T}{(2\pi)^3 \hbar \omega_{ho}}\int_{k_m({\bf x})}^{k_c({\bf x})} \!\!\! \frac{g(k,{\bf x}) k^2dk}{g(k,{\bf x})^2 - f({\bf x})^2} = 2(k_c({\bf x}) - k_m({\bf x})) \nonumber \\
&& \!\!\!\!\!\!\!\!\!\!\!\!\!\!\!\!\!\!\!\!\!\!\!\!\!\!\!\!\! - h^-({k_c,{\bf x}}) \arctan \left( \frac{k_c({\bf x})}{h^-({k_c,{\bf x}})} \right)
- h^+({k_c,{\bf x}}) \arctan \left( \frac{k_c({\bf x})}{h^+({k_c,{\bf x}})} \right) \nonumber \\
&& \!\!\!\!\!\!\!\!\!\!\!\!\!\!\!\!\!\!\!\!\!\!\!\!\!\!\!\!\! + h^-({k_m,{\bf x}}) \arctan \left( \frac{k_m({\bf x})}{h^-({k_m,{\bf x}})} \right)
+ h^+({k_m,{\bf x}}) \arctan \left( \frac{k_m({\bf x})}{h^+({k_m,{\bf x}})} \right), \label{eqn_nTRJ}
\end{eqnarray}
where $g(k,{\bf x}) = k^2/2+\tilde{V}_\text{ext}({\bf x})+2\gamma n_0({\bf x}) - \tilde{\mu}$, $f({\bf x}) = \gamma n_0({\bf x})$, and $h^+(k,{\bf x}) = [2g(k,{\bf x})+2f({\bf x})]^{1/2}$, $h^-(k,{\bf x}) = [2g(k,{\bf x})-2f({\bf x})]^{1/2}$.
Therefore, we have reduced the solution of the BdG system to a set of algebraic equations to be solved self-consistently. We note that the integral in Eq.\ (\ref{eqn_nTRJ}) diverges unless $k_c({\bf x})$ is finite and is the well known ultraviolet catastrophe that arises in a classical spectrum with equipartition of energy. The choice of $k_c({\bf x})$ is the single most important parameter that must be set in a $c$-fields simulation. We have set $k_c({\bf x})$ to correspond to values of $n_T$ of order unity. For $n_T\ll1$, the semi-classical approximation breaks down and an RJ distribution significantly deviates from the Bose-Einstein (BE) distribution. With this value of $n_T$, we have computed two cutoffs for $k_c$. The first (cutoff 1) was determined using the BdG expression for the energy in the semiclassical approximation \cite{Pethick02} with $E_c = 1.42$. A second cut-off using the same value of $E_c$ but with $k_c({\bf x})$ now computed from the expression of the energy for a single-particle in a harmonic trap was also used. This choice is motivated by the cut-offs typically used in $c$-field methods where the truncation is in terms of the single-particle basis functions. The values of $k_m({\bf x})$ in Eq.\ (\ref{eqn_nTRJ}) are determined from the BdG expression for the energy by specifying a minimum energy $E_m=0.01$ as a lower bound for the integral. In Fig.\ \ref{fig_cond}b, we present numerical results of the condensate and the thermal cloud densities obtained from a self-consistent solution of the BdG equations with a BE and a RJ distribution using the method described in \cite{Hutchinson97}, together with the analytical results obtained using the first order expression for the condensate density. We have compared our analytical solutions obtained using these cut-offs against numerical results computed for the BdG system using generalized Laguerre basis functions. Such a single-particle basis provides us with more control of how to truncate the spectrum in our computations and has become the hallmark of many $c$-field methods \cite{cfield}. In both cases, we see that the analytical solutions are in excellent agreement with the numerical results. As expected, the thermal cloud attains a maximum near the nonlinear turning point but since our approximation for the condensate is smooth, no cusp arises in the distribution of the thermal cloud.
\vspace{0.3cm} 

Our final system is concerned with that of a nonequilibrium condensate.
In recent years, there has been a surge of interest in modeling nonequilibrium systems such as exciton-polariton condensates \cite{Wouters07}. Recent work has shown that the steady state condensate density of these systems can be markedly different from the equilibrium Bose gases. In particular, we consider a 2D exciton-polariton condensate under conditions studied in \cite{Keeling08}. In that work, it was shown that for a given range of parameters, the 2D condensate profile remains radially symmetric in a harmonic trap. However, the presence of pumping and dissipation means that a radial flow is established from 
the edges of the condensate towards the center. The flow significantly modifies the profile of the condensate from that obtained with a TF approximation. Here, we show how the cubic WKB method can be used to approximate the condensate density in this nonequilibrium system.

We begin by considering $\mathcal{S} = \alpha \hbar \omega_{ho}$ and $\mathcal{D} = \sigma g|\psi|^2/\hbar \omega_{ho}$, as in \cite{Keeling08}, and seek a radially symmetric solution. Non-dimensionalising the cGPE equation in a similar way to Eq.\ (\ref{eqn_GP}), we see that a WKB ansatz applies provided $a_{ho} \ll R$. Proceeding as above in the case of the vortex solution, we seek a radially symmetric solution where 
$\psi(r)=U(x)= \exp{([\varphi(x)+i\theta(x)]/\epsilon)}$ with $r=\mathrm{e}^x$ to obtain two equations corresponding to the real and imaginary parts
\begin{eqnarray}
\mu \text{e}^{2x} &=& \frac{-\epsilon (\varphi'')}{2} - 
\frac{(\varphi'^2 - \theta'^2)}{2} + \frac{\text{e}^{4x}}{2} + \gamma \text{e}^
{2\varphi/\epsilon+2x},  \label{eqn_noneq1} \\
0 &=& \frac{-\epsilon (\theta'')}{2} - (\varphi' \theta') + (\alpha - \sigma \text{e}^{2\varphi/\epsilon})\text{e}^{2x}, \label{eqn_noneq2}
\end{eqnarray}
respectively. We now proceed as before by differentiating Eq.\ (\ref{eqn_noneq1}) and using it to simplify the resulting expression. This gives 
\begin{eqnarray} 
&&\varphi' \left\{ \varphi'^2 + \epsilon \varphi' -\theta'^2 + (2\mu\text{e}^{2x}-\text{e}^{4x})\right\} + \epsilon (\theta' \theta'' + \text{e}^{4x}-\theta'^2) \nonumber \\
&& = \epsilon^2 (\varphi'''/2-\varphi''). \nonumber \label{eqn_varphip_noneq}
\end{eqnarray}
In order to approximate this equation, let us consider the region near the origin where the exponential terms can be neglected. Moreover, the velocity at the origin must vanish since $\theta'(x) = \tilde{\theta}'(r)r$. We, can therefore rewrite the above equation in the vicinity of the origin in terms of $\tilde{\varphi}'(r)$ as 
\begin{eqnarray}
\tilde{\varphi}'^3 + \frac{\epsilon \tilde{\varphi}'^2}{r} = \epsilon^2 \left( \frac{\tilde{\varphi}'''}{2}+\frac{\tilde{\varphi}''}{2r}
-\frac{\tilde{\varphi}'}{2r^2} \right). \nonumber
\end{eqnarray}
Now we want to enforce the boundary condition $d\psi/dr = 0$ at the origin. However, from the above equation, we see that by neglecting all terms on the right-hand side, we can only enforce $\tilde{\varphi}'(0)=0$. In order to enforce $d\psi/dr = 0$, we must also have that $\tilde{\varphi}''(0)=0$. We can achieve this by retaining the last term proportional to $\tilde{\varphi}'(r)$ appearing on the right hand side. Our full approximate form of $\varphi'(x)$ is therefore given by
\begin{eqnarray}
&&\varphi' \left\{ \varphi'^2 + \epsilon \varphi' -\theta'^2 + \left( 2\mu\text{e}^{2x}-\text{e}^{4x} + \frac{\epsilon^2}{2} \right) \right\}  \nonumber \\
&& + \epsilon (\theta' \theta''+ \text{e}^{4x}-\theta'^2) = 0. \nonumber \label{eqn_varphip_noneq}
\end{eqnarray}

We can now solve for $\varphi'$ given $\theta'$ which has the solution 
\begin{eqnarray}
\theta'(x) = (2/\epsilon)\left[ \int_{-\infty}^{x}(\alpha-\sigma\text{e}^{2\varphi/\epsilon})\text{e}^{2\varphi/\epsilon+2x'}dx' \right]\text{e}^{-2\varphi/\epsilon}. 
\end{eqnarray}
This provides an expression for the flow which is a function of $\varphi(x)$. Hence, these equations provide solutions to our nonequilibrium system which are expressed in terms of a quadrature for the density $\text{e}^{2\int_{-\infty}^x \varphi'dx'}$ and the velocity $\theta'(x)=\tilde{\theta}'(r)$ and must be solved self-consistently. In practice, the above integrals are more easily evaluated by transforming back to $r$ space.

As before, the solution we have obtained specifies the wavefunction upto some normalization coefficient which we will denote by $C$. However, in the case of a nonequilibrium condensate, we are dealing with an open system with a nonconserved number of particles. Therefore, in order to compute the chemical potential we make use of the equation for $\theta'(x)$ and choose $C$ such that
\begin{eqnarray}
\int_{-\infty}^{\infty} (\alpha- C^2 \sigma\text{e}^{2\varphi/\epsilon})
\mathrm{e}^{2\varphi/\epsilon+2x'}dx' = 0. 
\end{eqnarray}
\vspace{0.1cm}
This equation defines $C$ which once computed can be used to evaluate the chemical potential with the aid of Eq.\ (\ref{eqn_noneq1}). By evaluating the terms in the equation at the origin, we find $\mu =  \gamma C^2$.

The solution of this coupled system obtained using only the leading order approximation for the condensate density with $\gamma = 0.5$, $\alpha = 2.2 \Theta(r_{\text{pump}}-r)$, and $\sigma = 0.15$, where $r_{\text{pump}}=4$ is the size of the pumping spot and $\Theta$ is the step function is shown in Fig.\ 1c. This corresponds to a chemical potential of $\mu = 10.97$ which agrees closely with the value of $11.18$ obtained from a full numerical solution of the cGPE equation.
This numerical solution of the radial cGPE computed with a pseudospectral method using Laguerre-basis functions is shown in Fig.\ 1c. As can be seen, the results are in excellent agreement even for this system in which the condensate density is significantly modified by the internal flow.

The author acknowledges partial support from the Isaac Newton Trust. The author would also like to thank Carlos Lobo, Magnus Borgh, and Natalia Berloff for many helpful discussions.

\end{document}